\documentclass{article}[9pt]
\usepackage{spconf,amsmath,epsfig}
\ninept

\title{Generalized Reduced-Rank Decompositions Using Switching and Adaptive Algorithms for Space-Time Adaptive
Processing
 \vspace{-0.15em}} \name{Rodrigo C. de Lamare
\vspace{-0.75em}}
\address{Communications Research Group, Department of
Electronics, \\
    University of York, United Kingdom \\
    Email: rcdl500@ohm.york.ac.uk \vspace{-0.35em}
    }
\begin{document}
\maketitle

\begin{abstract}
This work presents generalized low-rank signal decompositions with
the aid of switching techniques and adaptive algorithms, which do
not require eigen-decompositions, for space-time adaptive
processing. A generalized scheme is proposed to compute low-rank
signal decompositions by imposing suitable constraints on the
filtering and by performing iterations between the computed
subspace and the low-rank filter. An alternating optimization
strategy based on recursive least squares algorithms is presented
along with switching and iterations to cost-effectively compute
the bases of the decomposition and the low-rank filter. An
application to space-time interference suppression in DS-CDMA
systems is considered. Simulations show that the proposed scheme
and algorithms obtain significant gains in performance over
previously reported low-rank schemes. \vspace{-0.25em}
\end{abstract}

\vspace{-0.5em}

\section{Introduction}

Low-rank signal processing is an area that is central for dealing
with high-dimensional data, low-sample support situations and
large optimization problems that has gained considerable attention
in the last decades \cite{haykin,scharf2}. The origins of low-rank
modelling and signal processing lie in the problem of feature
selection encountered in statistical signal processing, which
refers to a dimensionality reduction process whereby a data space
is transformed into a feature space \cite{scharf2}. The
fundamental idea is to devise a decomposition that performs
dimensionality reduction so that the data vector can be
represented by a reduced number of effective features and yet
retain most of the intrinsic information content of the input data
\cite{scharf2}. The goal is to find an appropriate trade-off
between model bias and variance in a cost-effective way, yielding
a reconstruction error as small as desired.

Prior work has shown that low-rank adaptive filters
\cite{strobach}-\cite{delamaretsp09} are cost-effective techniques
for modelling a number of practical problems in acoustics,
communications, radar and sonar, and for dealing with large
filters and situations of short data records. Several low-rank
adaptive filtering methods have been proposed in the last decade
or so \cite{strobach}-\cite{delamaretsp09}. Among these techniques
are eigen-decomposition techniques \cite{strobach}, the multistage
Wiener filter (MSWF) \cite{goldstein}, the auxiliary vector
filtering (AVF) algorithm \cite{avf}, the interpolated
reduced-rank filters \cite{delamarespl1}, the reduced-rank filters
based on joint and iterative optimization (JIO)
\cite{delamarespl2} and joint iterative interpolation, decimation,
and filtering (JIDF) \cite{delamaretsp09}. Key problems with
previously reported low-rank adaptive schemes are the modelling of
certain low-rank signals and the design of multichannel processing
schemes \cite{strobach}-\cite{delamaretsp09}. Low-rank signals
that exhibit highly correlated statistical features and operate in
the presence of high-power signals (eg. jamming signals)
constitute a challenge for existing methods. Moreover, most
available methods require either separate structures for
multichannel processing \cite{delamaretsp09} or exhibit high
complexity \cite{avf,goldstein}.

In this work, a generalized scheme is devised to compute low-rank
signal decompositions with switching techniques and adaptive
algorithms, without the need for eigen-decompositions. The
proposed generalized low-rank decomposition with switching (GLRDS)
scheme computes the subspace and the low-rank filter that best
match the problem of interest with very fast convergence speed and
low complexity. By imposing constraints on the decomposition and
performing iterations between the computed subspace and the
low-rank filter, the proposed GLRDS scheme obtains smaller
reconstruction errors than existing methods. In order to compute
the parameters required in the signal decomposition and the
low-rank filter, an alternating optimization strategy based on
recursive least squares (RLS) algorithms is presented along with
switching and iterations to cost-effectively compute them. Unlike
existing schemes, the GLRDS efficiently lends itself to
multichannel processing. An application to space-time interference
suppression in DS-CDMA systems is considered. Simulations show
that the GLRDS scheme and algorithms obtain significant gains in
performance over existing schemes.

The paper is organized as follows. Section 2 formulates the
problem. Section 3 presents the proposed GLRDS scheme and least
squares (LS) design. Section 4 presents the alternating
optimization strategy along with recursive algorithms. Section 5
presents and discusses the simulation results and Section 6 draws
the conclusions.\vspace{-0.05em}

\section{Problem Statement}

In this section, the fundamental ideas of low-rank signal
processing are presented. The main design problems for a
decomposition that performs dimensionality reduction are
discussed. An approach based on linear algebra and a linear signal
model, which is sufficiently general to account for numerous
applications and topics, is adopted. Consider the following linear
signal model at time instant $i$ with $M$ samples organized in a
vector as given by
\begin{equation}
{\boldsymbol r}[i] = {\boldsymbol H} {\boldsymbol s}[i] +
{\boldsymbol n}[i],~~~ i=1,2, \ldots, P \label{obsig}
\end{equation}
where ${\boldsymbol r}[i]$ is the $M \times 1$ observed signal
vector with the $M$ samples to be processed, ${\boldsymbol H}$ is
the $M \times M$ matrix that describes the mixing nature of the
model, ${\boldsymbol s}[i]$ is the $M \times 1$ signal vector that
is generated by a given source, ${\boldsymbol n}[i]$ is an
$M\times 1$ vector of noise samples, and $P$ is the number of
observed signal vectors or the data record size.

\begin{figure}[!htb]
\begin{center}
\def\epsfsize#1#2{1\columnwidth}
\epsfbox{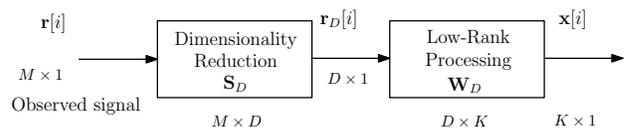} \vspace{-0.85em} \caption{Block diagram with
the stages of low-rank signal processing.}\vspace{-0.5em}
\label{fig1}
\end{center}
\end{figure}

In low-rank signal processing, the main idea is to process the
observed signal ${\boldsymbol r}[i]$ in two stages, as illustrated
in Fig. \ref{fig1}. The first stage corresponds to the
dimensionality reduction, whereas the second is responsible for
the signal processing in a lower-dimensional subspace. The
dimensionality reduction is performed by a mapping represented by
a decomposition matrix ${\boldsymbol S}_D = \big[{\boldsymbol s}_1
~\ldots ~ {\boldsymbol s}_{d} ~\ldots ~ {\boldsymbol s}_{D}\big]$
with dimensions $M \times D$, where $D$ is the rank ($D < M$) that
projects ${\boldsymbol r}[i]$ onto a $D \times 1$
reduced-dimension data vector ${\boldsymbol r}_D[i]$ and
${\boldsymbol s}_{d}$ is the $d$th column of ${\boldsymbol S}_D$.
This relationship is expressed by
\begin{equation}
\begin{split}
{\boldsymbol r}_D[i] & = {\boldsymbol S}_D^H {\boldsymbol r}[i]  =
\sum_{d=1}^{D} {\boldsymbol s}_d^H {\boldsymbol r}[i] {\boldsymbol
q}_d ,
\end{split}
\end{equation}
where ${\boldsymbol q}_d$ is a $D \times 1$ vector with a one in
the $d$th position and zeros elsewhere, $(\cdot)^{T}$ and
$(\cdot)^{H}$ denote transpose and Hermitian transpose,
respectively. Key design criteria for the matrix ${\boldsymbol
S}_D$ and the dimensionality reduction are the reconstruction
error, the computational complexity and the compression ratio
${\rm CR}=M/D$. These parameters usually depend on the application
and the design requirements.

After the dimensionality reduction, an algorithm is used to
perform the signal processing task on the reduced-dimension
observed vector ${\boldsymbol r}_D[i]$ according to the designer's
aims. The resulting scheme with $D$ elements shall benefit from a
reduced number of parameters, which may lead to lower complexity,
smaller requirements for storage, faster convergence and superior
tracking capability. In the case of filtering by a $D \times K$
matrix ${\boldsymbol W}_D =[ {\boldsymbol w}_{D,1} ~{\boldsymbol
w}_{D,2}~\ldots {\boldsymbol w}_{D,1}]$, we have the following
output $K \times 1$ vector estimate
\begin{equation}
\begin{split}
\hat{\boldsymbol x}[i] & = {\boldsymbol W}^{H}_D {\boldsymbol
S}_D^H {\boldsymbol r}[i] = {\boldsymbol W}^{H}_D \sum_{d=1}^{D}
{\boldsymbol s}_d^H {\boldsymbol r}[i] {\boldsymbol q}_d\\
& = \sum_{k=1}^{K} {\boldsymbol w}^H_{D,k} \big( \sum_{d=1}^{D}
{\boldsymbol s}_d^H {\boldsymbol r}[i] {\boldsymbol q}_d \big)
{\boldsymbol q}_k   .
\end{split}
\end{equation}
We consider low-rank algorithms with the aid of linear design
techniques. In order to process ${\boldsymbol r}[i]$ with low-rank
techniques, we need to solve the mean-square error (MSE)-based
optimization problem
\begin{equation}
\begin{split}
\hspace{-1.05em} \big[{\boldsymbol S}_{D, {\rm opt}}, {\boldsymbol W}_{D,{\rm opt}}
\big] & = \arg \min_{ {\boldsymbol S}_{d}, {\boldsymbol W}_D}
E\big[ || {\boldsymbol x}[i] - {\boldsymbol W}^{H}_D{\boldsymbol
S}_{D}^H{\boldsymbol r}[i]||^2 \big], 
\label{ms1}
\end{split}
\end{equation}
where ${\boldsymbol x}[i]$ is the desired signal and $E[\cdot]$
stands for the expected value operator. The optimal solution
${\boldsymbol W}_{D,{\rm opt}}$ of the problem in (\ref{ms1}) is
obtained by fixing ${\boldsymbol S}_D$, taking the gradient terms
of the argument with respect to ${\boldsymbol W}_D^*$ and equating
them to a zero matrix which yields
\begin{equation}
\begin{split}
\hspace{-1.25em}{\boldsymbol W}_{D,{\rm opt}} & = \bar{\boldsymbol
R}^{-1}\bar{\boldsymbol P} = \big({\boldsymbol S}_D^H{\boldsymbol
R}{\boldsymbol S}_D\big)^{-1}{\boldsymbol S}_D^H{\boldsymbol P},
\label{opv}
\end{split}
\end{equation}
where $\bar{\boldsymbol R} = E\big[\bar{\boldsymbol
r}[i]\bar{\boldsymbol r}^{H}[i]\big]={\boldsymbol
S}_D^H{\boldsymbol R}{\boldsymbol S}_D$ is the $D \times D$
low-rank correlation matrix, ${\boldsymbol R} = E\big[{\boldsymbol
r}[i]{\boldsymbol r}^{H}[i]\big]$ is the $M \times M$
full-rank correlation matrix, $\bar{\boldsymbol P}= E\big[
\bar{\boldsymbol r}[i]{\boldsymbol x}^H[i]\big]={\boldsymbol
S}_D^H{\boldsymbol P}$ is the $D \times K$ cross-correlation
matrix of the low-rank model. The associated MMSE for a rank-$D$
matrix filter is expressed by
\begin{equation}
\begin{split}
{\rm MMSE} & =  
\sigma^2_x - {\rm tr}\big[{\boldsymbol P}^H{\boldsymbol
S}_D ({\boldsymbol S}_D^H{\boldsymbol R}{\boldsymbol S}_D)^{-1}
{\boldsymbol S}_D^H{\boldsymbol P}\big], \label{mse}
\end{split}
\end{equation}
where $\sigma^2_x=E\big[| {\boldsymbol x}^H[i]{\boldsymbol x}[i]
|^2\big]$ and ${\rm tr}[ \cdot ]$ stands for trace. The optimal
solution ${\boldsymbol S}_{D,{\rm opt}}$ of the problem in
(\ref{ms1}) is obtained by fixing ${\boldsymbol w}_D[i]$, taking
the gradient terms of the associated MMSE in (\ref{mse}) with
respect to ${\boldsymbol S}_D^*$ and equating them to a zero
matrix. Considering the eigen-decomposition of ${\boldsymbol R} =
{\boldsymbol \Phi} {\boldsymbol \Lambda} {\boldsymbol \Phi}^H$,
where ${\boldsymbol \Phi}$ is an $M \times M$ unitary matrix with
the eigenvectors of ${\boldsymbol R}$ and ${\boldsymbol \Lambda}$
is an $M \times M$ diagonal matrix with the eigenvalues of
${\boldsymbol R}$ in decreasing order, we have
\begin{equation}
{\boldsymbol S}_{D,{\rm opt}} = {\boldsymbol \Phi}_{1:M,1:D},
\end{equation}
where ${\boldsymbol \Phi}_{1:M,1:D}$ is a $M\times D$ unitary
matrix that corresponds to the signal subspace and contains the
$D$ eigenvectors associated with the $D$ largest eigenvalues of
the unitary matrix ${\boldsymbol \Phi}$. In our notation, the
subscript represents the number of components in each dimension.
For example, the $M\times D$ matrix ${\boldsymbol \Phi}_{1:M,1:D}$
contains the $D$ first columns of ${\boldsymbol \Phi}$, where each
column has $M$ elements.

The previous development suggests that the central element for
constructing low-rank techniques is the design of ${\boldsymbol
S}_D$ since the MMSE in (\ref{mse}) depends on ${\boldsymbol P}$,
${\boldsymbol R}$ and ${\boldsymbol S}_D$. The quantities
${\boldsymbol P}$ and ${\boldsymbol R}$ are common to both
low-rank and full-rank designs, however, the matrix ${\boldsymbol
S}_D$ plays a key role in the dimensionality reduction and in the
performance. In what follows, a cost-effective scheme for
computing ${\boldsymbol S}_D$, ${\boldsymbol W}_{D,{\rm opt}}$ and
the remaining statistical quantities is presented. \vspace{-0.35em}

\section{Proposed GLRDS Scheme and LS Design}
\vspace{-0.85em}
\begin{figure}[!htb]
\begin{center}
\def\epsfsize#1#2{1\columnwidth}
\epsfbox{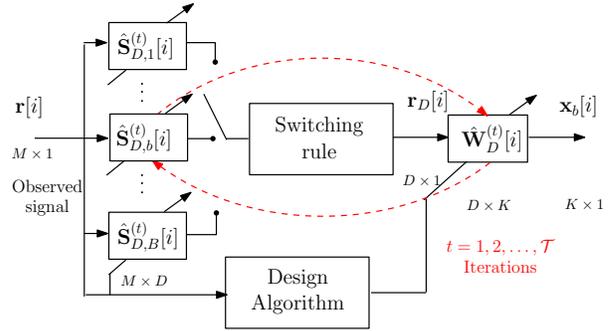} \vspace{-0.95em} \caption{Proposed GLRDS
scheme.}\vspace{-0.5em} \label{fig2}
\end{center}
\end{figure}

The idea of the proposed GLRDS scheme, shown in Fig. 2, is to
introduce constraints in the decomposition matrix ${\boldsymbol
S}_D$ to substantially reduce the number of parameters for
filtering. Since this usually affects the reconstruction error of
the algorithm, a switching mechanism is incorporated to provide
the ${\boldsymbol S}_D$ matrix with alternative bases. Similar
ideas have been reported in the literature of automatic control
and, more specifically, in the area of switched control techniques
\cite{liberzon,xuping}. Consider the following vector estimate:
\begin{equation}
\begin{split}
\hat{\boldsymbol x}_b[i] & = {\boldsymbol W}^{H}_D[i] {\boldsymbol
S}_{D,b}^H[i] {\boldsymbol r}[i]  \\
& = {\boldsymbol W}^{H}_D[i] \bigg( \sum_{d=1}^{D} {\boldsymbol
q}_d {\boldsymbol d}^H_{d,b} {\boldsymbol C}_{{\boldsymbol
s}_{d,b}}[i] \bigg) {\boldsymbol r}[i]
\\ & =  {\boldsymbol W}^{H}_D[i] \bigg( \sum_{d=1}^{D}
{\boldsymbol q}_d {\boldsymbol d}^H_{d,b} {\boldsymbol
C}_{{\boldsymbol r}}[i] \bigg) {\boldsymbol s}_{d,b} [i] ,
\label{vest}
\end{split}
\end{equation}
where ${\boldsymbol d}_{d,b}[i]$ is the $M\times 1$ shaping vector
employed to mould the $d$th column of the matrix ${\boldsymbol
S}_{D,b}^H[i]$ which is expressed as
\begin{equation}
{\boldsymbol d}_{d,b}[i] = [\underbrace{0~~ \ldots ~~
0}_{\gamma_{d}~zeros} ~~~ 1 ~~  \underbrace{ 0  ~~~ \ldots ~~
0}_{(M-\gamma_{j}-1) ~zeros} ]^T, \label{decvec}
\end{equation}
and $b=1,~\ldots~, B$ is the index of parallel switching branches.
The quantity $\gamma_{j}$ is the number of zeros chosen according
to a given design criterion. In this work, we use patterns created
by $\gamma_{j} = (j-1) \lfloor M/D \rfloor +(b-1)$ due to their
simplicity and satisfactory performance. These patterns lead to
$I_d \times 1$ basis vectors ${\boldsymbol s}_{d,b}[i]$. The $M
\times I_d$ matrices ${\boldsymbol C}_{{\boldsymbol s}_{d,b}}[i]$
and ${\boldsymbol C}_{{\boldsymbol r}}[i]$ are Hankel matrices
with shifted versions of ${\boldsymbol s}_{d,b}[i]$ and
${\boldsymbol r}[i]$ described by
\begin{equation}
{\boldsymbol C}_{\boldsymbol r}[i]=\left[\begin{array}{c c c c c}
r_{0}^{[i]} & r_{1}^{[i]}  & \ldots & r_{I_d-1}^{[i]}  \\
r_{1}^{[i]}  & r_{2}^{[i]} & \ldots & r_{I_d}^{[i]}  \\
\vdots & \vdots  & \ddots & \vdots \\
r_{M-2}^{[i]}  & r_{M-1}^{[i]}  & \ldots & 0  \\
r_{M-1}^{[i]}  & 0  & \ldots & 0  \\
 \end{array}\right]. \label{hankel}
\end{equation}
A similar structure to the above can be obtained for ${\boldsymbol
C}_{{\boldsymbol s}_{d,b}}[i]$. In order to design the GLRDS
scheme, the parameters $I_d$, $D$ and $B$ must be chosen, and the
filters ${\boldsymbol s}_{d,b}[i]$ and ${\boldsymbol W}_D[i]$ need
to be computed by solving the optimization problem
\begin{equation}
\begin{split}
\big[{\boldsymbol s}_{d,b}^{\rm opt}, {\boldsymbol W}_{D}^{\rm
opt} \big] & = \arg \min_{ {\boldsymbol s}_{d,b}[i], {\boldsymbol
W}_D[i]} \sum_{l=1}^i \lambda^{i-l} || {\boldsymbol x}[l] -
\hat{\boldsymbol x}_b[l]||^2 \big],
\\ d  &=1,~ \ldots, ~D, ~ b=1,~ \ldots, ~ B. \label{design}
\end{split}
\end{equation}
Fixing ${\boldsymbol W}_D[i]$ and solving the problem for
${\boldsymbol s}_{d,b}[i]$, we obtain
\begin{equation}
\begin{split}
{\boldsymbol s}_{d,b}[i] & =  {\boldsymbol R}_{d,b}^{-1}[i]\big(
{\boldsymbol p}_{d,b}[i] - \sum_{j\neq d}^{D} {\boldsymbol
P}_{j,b}[i] {\boldsymbol s}_{j,b}[i] \big),
\\ d,j &  =1,~ \ldots, ~D,~ ~ b=1,~ \ldots, ~ B, \label{svec}
\end{split}
\end{equation}
where  ${\boldsymbol R}_{d,b}[i] = \sum_{l=1}^{i} \lambda^{i-l}
{\boldsymbol q}_d^H {\boldsymbol W}_D[i]{\boldsymbol
W}^H_D[i]{\boldsymbol q}_d {\boldsymbol C}_{\boldsymbol r}^T[l]
{\boldsymbol d}_{d,b} {\boldsymbol d}_{d,b}^H {\boldsymbol
C}_{\boldsymbol r}^*[l]$  and ${\boldsymbol P}_{j,b}[i] =
\sum_{l=1}^{i} \lambda^{i-l} {\boldsymbol q}_j^H {\boldsymbol
W}_D[i]{\boldsymbol W}^H_D[i]{\boldsymbol q}_d {\boldsymbol
C}_{\boldsymbol r}^T[l] {\boldsymbol d}_{d,b} {\boldsymbol
d}_{j,b}^H {\boldsymbol C}_{\boldsymbol r}^*[l]$ are $I_d \times
I_d$ correlation matrices, and the $I_d \times 1$ vector
${\boldsymbol p}_{d,b}[i]= \sum_{l=1}^{i}\lambda^{i-l}{\boldsymbol
x}^H[l] {\boldsymbol W}^H_D[i]{\boldsymbol q}_d {\boldsymbol
C}_{\boldsymbol r}^T[l] {\boldsymbol d}_{d,b} $ is a
cross-correlation vector.

Once the ${\boldsymbol s}_{d,b}[i]$ are computed, we can build the
corresponding decomposition matrix ${\boldsymbol S}_{D,b}[i]$.
According to the schematic in Fig. \ref{fig2}, a selection is
performed among the $B$ available matrices as follows
\begin{equation}
{\boldsymbol S}_{D,b}[i] = {\boldsymbol S}_{D,b_{\rm s}}[i] ~~
\textrm{when} ~~ b_{\rm s} = \arg \min_{1\leq b \leq B}
||\underbrace{{\boldsymbol x}[i] - \hat{\boldsymbol
x}_{b}[i]}_{{\boldsymbol e}_b[i]}||^{2}, \label{SDdesign}
\end{equation}
Now fixing ${\boldsymbol s}_{d,b}[i]$ and solving the problem for
${\boldsymbol W}_D[i]$, we have
\begin{equation}
{\boldsymbol W}_D[i+1] = {\boldsymbol R}^{-1}[i]{\boldsymbol
P}[i],\label{wmat}
\end{equation}
where the matrix ${\boldsymbol R}[i] = \sum_{l=1}^{i}
\lambda^{i-l} {\boldsymbol S}_{d,b}^{H^{(t)}}[l] {\boldsymbol
r}[l]{\boldsymbol r}^H[l] {\boldsymbol S}_{d,b}^{(t)}[l]$ is a $D
\times D$ correlation matrix and ${\boldsymbol P}^{(t)}[i] =
\sum_{l=1}^{i} \lambda^{i-l} {\boldsymbol S}_{d,b}^{H^{(t)}}[l]
{\boldsymbol r}[l] {\boldsymbol x}^H[l]$ is a $D \times K$
cross-correlation matrix. The LS algorithm outlined in
(\ref{design})-(\ref{wmat}) can be efficiently computed in a
recursive fashion with alternating steps, as will be shown in the
next section. \vspace{-0.35em}

\section{Proposed Adaptive Algorithms} \vspace{-0.35em}

In this section, we present recursive alternating least squares
(RALS) algorithms. The basic idea of the RALS is to solve the LS
expressions in (\ref{svec})-(\ref{wmat}) via an alternating
strategy with $t=1, \ldots, {\mathcal T}$ iterations. Using the
expressions in (\ref{svec}) and the matrix inversion lemma
\cite{haykin}, we obtain for $d,j  =1,~ \ldots, ~D,~ b=1,~ \ldots,
~ B$, and $t=1, \ldots, {\mathcal T}$
\begin{equation}
\begin{split}
{\boldsymbol s}_{d,b}^{(t)}[i] & =  {\boldsymbol P}_{d,b}[i]\big(
{\boldsymbol p}_{d,b}[i] - \sum_{j\neq d}^{D} {\boldsymbol
P}_{j,b}[i] {\boldsymbol s}_{j,b}^{(t)}[i] \big), \label{svecr}
\end{split}
\end{equation}
where
\begin{equation}
{\boldsymbol P}_{d,b}[i] = \lambda^{-1} {\boldsymbol P}_{d,b}[i-1]
- \lambda^{-1} {\boldsymbol k}_{d,b}[i] {\boldsymbol
C}_{\boldsymbol r}^T[i] {\boldsymbol d}_{d,b} {\boldsymbol
P}_{d,b}[i-1],
\end{equation}
\begin{equation}
{\boldsymbol k}_{d,b}[i] = \frac{\lambda^{-1} {\boldsymbol
P}_{d,b}[i-1] {\boldsymbol d}_{d,b}^H {\boldsymbol C}_{\boldsymbol
r}^*[i]}{(\sum_{k=1}^{K} |w_{d,k}[i]|^2)^{-1} + \lambda^{-1}
{\boldsymbol d}_{d,b}^H {\boldsymbol C}_{\boldsymbol
r}^*[i]{\boldsymbol P}_{d,b}[i-1]{\boldsymbol C}_{\boldsymbol
r}^T[i] {\boldsymbol d}_{d,b}},
\end{equation}
\begin{equation}
{\boldsymbol P}_{j,b}[i] = \lambda^{-1} {\boldsymbol P}_{j,b}[i-1]
+ {\boldsymbol q}_j^H {\boldsymbol W}_D[i]{\boldsymbol
W}^H_D[i]{\boldsymbol q}_d {\boldsymbol C}_{\boldsymbol r}^T[i]
{\boldsymbol d}_{d,b} {\boldsymbol d}_{j,b}^H {\boldsymbol
C}_{\boldsymbol r}^*[i],
\end{equation}
\begin{equation}
{\boldsymbol p}_{d,b}[i]= \lambda {\boldsymbol p}_{d,b}[i-1] +
{\boldsymbol x}^H[i] {\boldsymbol W}^H_D[i]{\boldsymbol q}_d
{\boldsymbol C}_{\boldsymbol r}^T[i] {\boldsymbol d}_{d,b}
\end{equation}
With the ${\boldsymbol s}_{d,b}^{(t)}[i]$, we can build the
corresponding decomposition matrix ${\boldsymbol
S}_{D,b}^{(t)}[i]$ and perform the pattern selection as follows
\begin{equation}
{\boldsymbol S}_{D,b}^{(t)}[i] = {\boldsymbol S}_{D,b_{\rm
s}}^{(t)}[i] ~~ \textrm{when} ~~ b_{\rm s} = \arg \min_{1\leq b
\leq B} ||{\boldsymbol x}[i] - \hat{\boldsymbol
x}_{b}^{(t)}[i]||^{2}, \label{SDdesign2}
\end{equation}
After the selection of the decomposition matrix, we can construct
${\boldsymbol r}_D[i] = {\boldsymbol S}_{D,b_{\rm
s}}^{H^{(t=1)}}[i] {\boldsymbol r}[i]$ and compute the filter
${\boldsymbol W}_D^{(t)}[i+1]$
\begin{equation}
{\boldsymbol W}_D^{(t)}[i+1] = {\boldsymbol W}_D[i] + {\boldsymbol
k}_D[i] {\boldsymbol e}^{H^{(t)}}[i] ,\label{wmatr}
\end{equation}
where ${\boldsymbol e}^{(t)}[i] = {\boldsymbol x}[i] -
\hat{\boldsymbol x}_{b_{\rm s}}^{(t)}[i]$ and
\begin{equation}
{\boldsymbol k}_{D}[i] = \frac{\lambda^{-1} {\boldsymbol
P}_{D}[i-1] {\boldsymbol r}_{D}[i]}{(1 + \lambda^{-1} {\boldsymbol
r}_{D}^H[i]  {\boldsymbol P}_{D}[i-1]{\boldsymbol r}_{D}[i]},
\end{equation}
\begin{equation}
{\boldsymbol P}_{D}[i] = \lambda^{-1} {\boldsymbol P}_{D}[i-1] -
\lambda^{-1}  {\boldsymbol k}_{D}[i]{\boldsymbol
r}_{D}^H[i]{\boldsymbol P}_{D}[i-1], \label{milP}
\end{equation}
The proposed RALS algorithm consists of using
(\ref{svecr})-(\ref{milP}) with $t=1,\ldots, {\mathcal T}$
alternating iterations between the filters ${\boldsymbol
s}_{d,b}^{(t)}[i]$ and ${\boldsymbol W}_D^{(t)}[i]$. This allows a
very fast convergence for the GLRDS scheme and a significant
reduction of the MSE. To this end, we only need to iterate
(\ref{svecr}), the error ${\boldsymbol e}^{(t)}[i]$ and
(\ref{wmatr}). The complexity of the proposed GLRDS with the RALS
algorithm is $O(D^2)$ to compute ${\boldsymbol W}_D^{(t)}[i]$ and
$O(D(I_d^2))$ to compute ${\boldsymbol s}_{d,b}^{(t)}[i]$. Since
$I_d$ is typically very small ($I_d=2,3$) and the maximum number
of iterations ${\mathcal T}=2,3$ the complexity of the GLRDS with
the RALS algorithm is significantly lower than the full-rank RLS
\cite{haykin}, the eigen-decomposition methods \cite{strobach},
the MSWF \cite{goldstein}, and the AVF \cite{avf}. \vspace{-0.35em}

\section{Simulations} \vspace{-0.35em}

The performance of the GLRDS scheme and RALS algorithms is
assessed via simulations for space-time interference suppression.
We consider the uplink of a DS-CDMA system with symbol interval
$T$, chip period $T_c$, $QPSK$ modulation, spreading gain
$N=T/Tc$, $K$ users,  and equipped with  a uniform antenna array
with $J$ elements. The spacing between the antenna elements is
$d=\lambda_c/2$, where $\lambda_c$ is the carrier wavelength.
Assuming that the channel is constant during each symbol and the
receiver is synchronized with the main path, the received signal
after filtering by a chip-pulse matched filter and sampled at chip
rate yields the $M\times 1$ received vector \vspace{-0.35em}
\begin{equation}
\begin{split}
{\bf r}[i] & =  \sum_{k=1}^{K} A_{k}x_{k}[i] {\boldsymbol
p}_{k}[i] + {\boldsymbol \eta}[i] + {\boldsymbol j}[i]+
{\boldsymbol n}[i],
\end{split}
\end{equation}
where $M=J(N+L_p-1)$, the complex Gaussian noise vector is ${\bf
n}[i] = [n_{1}[i] ~\ldots~n_{M}[i]]^{T}$ with $E[{\bf n}[i]{\bf
n}^{H}[i]] = \sigma^{2}{\bf I}$. The $JL_p \times 1$ channel
vector ${\boldsymbol h}_{k}[i]$ contains the complex gains of the
channel from user $k$ to each antenna element. The $M\times 1$
spatial signature for user $k$ is ${\bf p}_{k}[i] = {\boldsymbol
{\mathcal F}}_{k} {\boldsymbol h}_{k}[i]$, where
${\boldsymbol{\mathcal F}}_{k}$ is an $M \times JL_p$ matrix with
shifted versions of the signature sequence ${\bf s}_{k} =
[a_{k}(1) \ldots a_{k}(N)]^{T}$ of user $k$ that performs
convolution of the channel ${\boldsymbol h}_{k}[i]$ with ${\bf
s}_{k}$. The signatures are randomly generated with $N=16$.  For
the simulations, we use the initial values ${\boldsymbol W}_D[0]=
[ {\bf 1}_{K \times 1} {\bf 0}_{K \times D-1} ]^T$ and
${\boldsymbol S}_D[0]=[{\bf I}_D ~ {\bf 0}_{D \times M-D}]^T$,
assume $L=9$ as an upper bound, use $3$-path channels with
relative powers given by $0$, $-3$ and $-6$ dB, where in each run
the spacing between paths is obtained from a discrete uniform
random variable between $1$ and $2$ chips and average the
experiments over $200$ runs. The power and the phase of each path
is time-varying and follows Clarke's model \cite{rappa}. The
system has a power distribution among the users for each run that
follows a log-normal distribution with associated standard
deviation equal to $1.5$ dB and there is a sinusoidal jamming
signal ${\boldsymbol j}[i]$ with a power level $20$ dB above the
average signal-to-noise ratio (SNR) of the users, which are
jointly demodulated.

\begin{figure}[!htb]
\begin{center}
\def\epsfsize#1#2{0.95\columnwidth}
\epsfbox{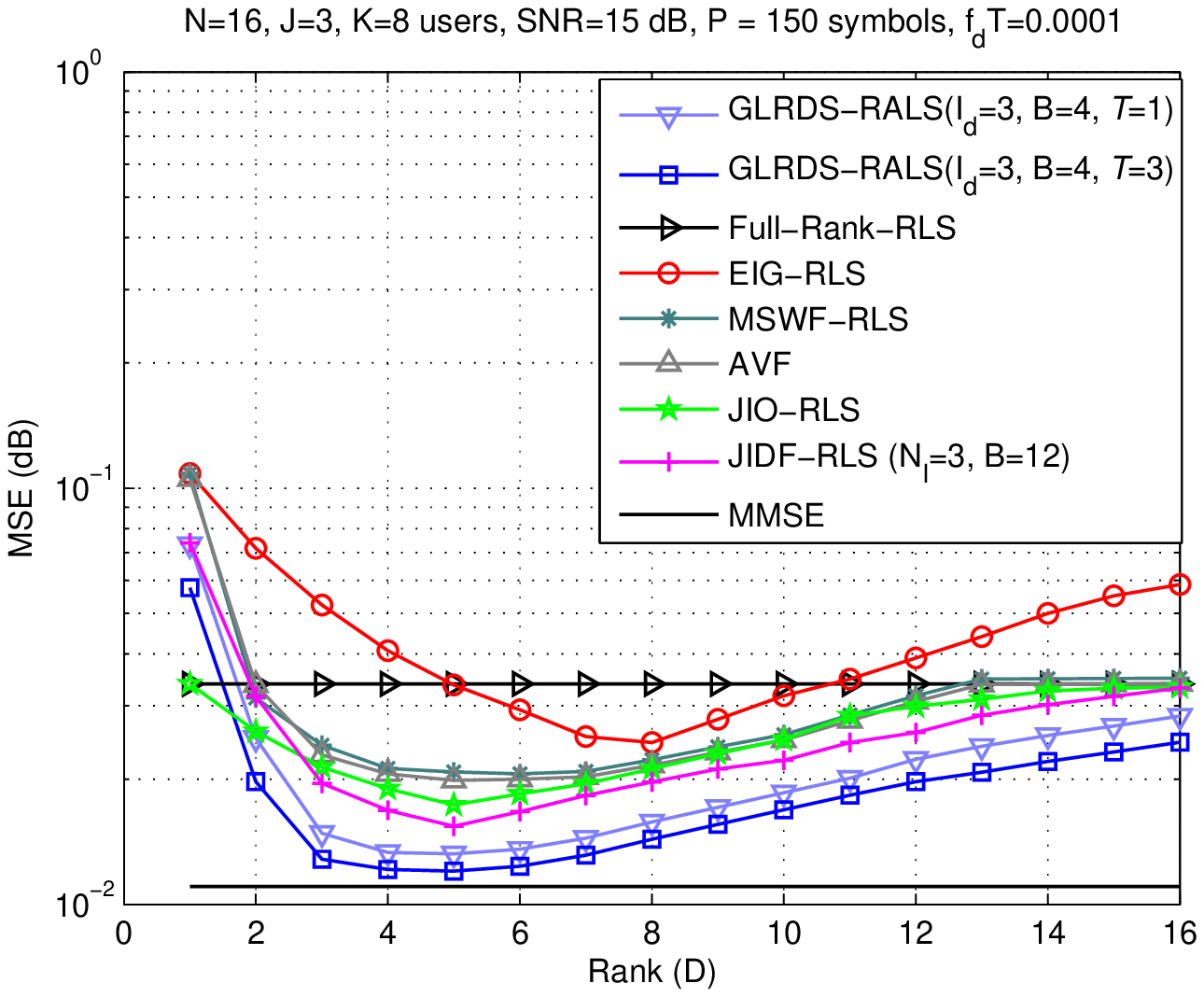} \vspace{-1.8em}\caption{\footnotesize MSE
performance versus rank (D). Parameters: $\lambda=0.999$,
${\boldsymbol P}_D[0]= 0.01 {\bf I}$, ${\boldsymbol P}_{d,b}[0]=
0.01 {\bf I}$.} \vspace{-1.5em}\label{fig3}
\end{center}
\end{figure}

We compare the proposed GLRDS scheme and RALS with the Full-rank
RLS \cite{haykin}, the eigen-decomposition (EIG) \cite{strobach},
the MSWF \cite{goldstein}, the AVF \cite{avf}, the JIO
\cite{delamarespl2} and JIDF \cite{delamaretsp09} techniques. We
consider the MSE performance versus the rank $D$ space-time
receivers process ${\boldsymbol r}[i]$ with $M=75$ samples per
symbol. The results in Fig. \ref{fig3} show that the best rank for
the GLRDS scheme is $D=4$ (used in the next experiments) and that it is
very close to the optimal MMSE. The results in Fig. \ref{fig3}
show that the best rank for the proposed scheme is $D=4$ and it is
very close to the optimal MMSE. The number of elements $I_d$
required to construct the decompositions is often small ( a
uniform $I_d=3$ for $d=1,\ldots, D$ is used here throughout) and a
designer can employ non-uniform lengths according the low-rank
modelling needs. The number of iterations ${\mathcal T}$ is also
typically small and allows the RALS to converge faster. Our
studies with systems with different sizes suggest that $D$ is
relatively invariant to the system size, which brings considerable
computational savings to the GLRDS scheme and allows a suitable
low-rank modelling and a very fast convergence performance. In
practice, the rank $D$ can be adapted in order to account for
time-varying scenarios and models, ensuring good performance and
tracking after convergence.

\begin{figure}[!htb]
\begin{center}
\def\epsfsize#1#2{0.95\columnwidth}
\epsfbox{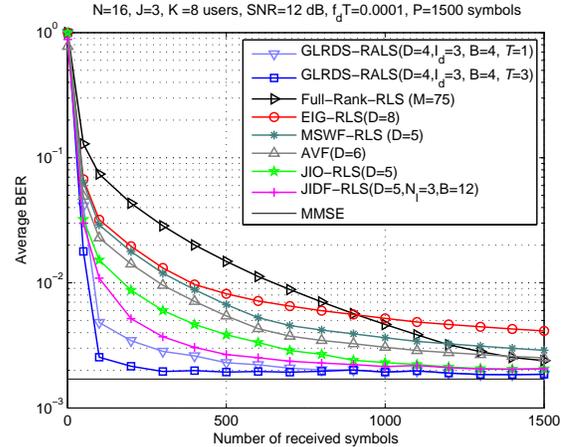} \vspace{-1.8em}\caption{\footnotesize BER
performance versus number of received symbols. Parameters:
$\lambda=0.999$, ${\boldsymbol P}_D[0]= 0.01 {\bf I}$, ${\boldsymbol
P}_{d,b}[0]= 0.01 {\bf I}$.} \vspace{-1.5em} \label{fig4}
\end{center}
\end{figure}

In the next experiment we evaluate the average BER performance
against the number of received symbols for the GLRDS and RALS, and
the existing schemes and algorithms, as depicted in Fig.
\ref{fig4}. The packet size is $P=1500$ symbols and the adaptive
filters are trained with $200$ symbols and then are switched to
decision-directed mode to continue the adaptation.The results show
that the proposed GLRDS scheme has a much better performance than
the existing approaches and is able to adequately track the
desired signal. A stability and convergence analysis of the
proposed scheme based on control-theoretic arguments
\cite{liberzon,xuping}, including tracking and steady-state
performance, conditions and proofs are not included here due to
lack of space and are intended for a future paper.

 \vspace{-0.35em}

\section{Conclusions}
 \vspace{-0.35em}
This work has proposed the GLRDS scheme and the RALS algorithms
for performing low-rank adaptive filtering, and has considered
their application to space-time interference suppression in
DS-CDMA systems. The GRLDS scheme provides a way for computing
generalized low-rank signal decompositions with switching
techniques and adaptive algorithms, which does not require
eigen-decompositions. The results of simulations show that the
proposed GRLDS scheme and the RALS algorithms obtain quite
significant gains in performance over previously reported low-rank
schemes.

\end{document}